\documentclass[a4paper,12pt]{article}
\usepackage{epsf,multicol,ifthen}
\usepackage[dvips]{graphicx}
\usepackage{cite}
\usepackage{hyperref}
\newcommand{\bec}{\begin{center}}
\newcommand{\ec}{\end{center}}
\newcommand{\bee}{\begin{equation}}
\newcommand{\ee}{\end{equation}}

\title{Mass reconstruction of MSSM Higgs boson}

\author{T.V. Obikhod\thanks{E-mail: obikhod@kinr.kiev.ua}, I.A. Petrenko\\
\small\emph{Institute for Nuclear Research, National Academy of Science of Ukraine} \\ 
\small\emph{47, prosp. Nauki, Kiev, 03028, Ukraine}}

\begin{document}
\maketitle

\abstract{The problems of Standard Model as well as questions related to Higgs boson properties led to the need for modeling  of ttH associated production and Higgs boson decay to top quark pair within the MSSM model. With the help of computer programs MadGraph, Pythia and Delphes and using the latest kinematic cuts taken from experimental data obtained at the LHC we predicted the masses of MSSM Higgs bosons, A and H. 
} \\ 
\vspace*{3mm}\\
{\bf Key words}: MSSM Higgs boson $\cdot$ top quark $\cdot$ b-tagging $\cdot$ computer modeling $\cdot$ the mass of Higgs boson.

\newpage
\section{Introduction}

	The study of the properties of the Higgs boson discovered in 2012 is one of the main objectives of the LHC, \cite{1.}. The importance of the experiments is related to the refinement of the channels of formation and decay of the Higgs boson, which show that there are deviations of more than 2$\sigma$ from the Standard Model (SM). Such data, together with the theoretical predictions of new physics, such as supersymmetry and the theory of extra dimensions, lead to the need to model the properties of the Higgs boson beyond the SM (BSM) such as production cross sections, angular distributions and masses of supersymmetric Higgs bosons.

	The existence of SM problems related to the impossibility of combining gravity with the other three types of interactions, the problem of radiative corrections to the Higgs boson mass, neutrino oscillations and dark matter and dark energy problems lead to the introduction of new theories, one of which is supersymmetry. There are many supersymmetric theories. We will further use Minimal Supersymmetric Standard Model (MSSM) for the prediction of new supersymmetric particles - superpartners of the Higgs boson. The advantage of such a search lies not only in the possibility of going beyond the framework of the SM, but also in the small mass of the Higgs superparticles provided for by the new theories. Such searches could be implemented both at the existing LHC collider, and at future accelerators of the type ILC or FCC. To establish a deviation from the SM behavior, the next goal is to identify the nature of electro-weak symmetry breaking (EWSB), which is connected with properties of top quark and Higgs boson interactions. Predictions for the coupling of the Higgs boson to top quarks directly influence on the measurements of the production and decay rates and angular correlations. Therefore, this information can be used to study probe whether data are compatible with the SM predictions for the Higgs boson. Since the QCD and electroweak gauge interactions of top quarks have been well established, the top Yukawa coupling might differ from the SM value. Therefore, the measurement of the ttH production rate and tt decay of A boson can provide direct information on the top-Higgs Yukawa coupling, probably the most crucial coupling to fermions. The anomalous interaction of the Higgs boson with the top quark, has been experimentally studied through the measurement of the Higgs boson production in association with a top quark, \cite{2.}. According to the combined analysis of the experimental data at the LHC, the constrain on the top quark Yukawa coupling, $y_t$, within [−0.9,\ −0.5] and [1.0,\ 2.1] $\times$ $y^{SM}_t$ were obtained. Recent ATLAS Higgs results using Run-2 data at a center-of-mass energy of 13 TeV with up to an integrated luminosity of 80 fb$^{-1}$ to probe BSM coupling for tH+ttH processes, \cite{3.} showed that Higgs boson will continue to provide an important probe for new physics and beyond. 
    
    To implement the searches for the MSSM Higgs bosons and to facilitate their findings, we chose a specific search channels and the methods by which the corresponding superparticles were identified. Using latest experimental data for ttH production of Higgs boson \cite{4.}, b-tagging algorithm and MadGraph, Pythia, Delphes programs and latest kinematic cuts we predicted the masses of superparticles, A and H. 

\section{B-tagging identification and reconstruction of MSSM Higgs boson masses}    
    Top-quark Yukawa coupling $y_t$ is parameterized as following
\[L_{Htt}=-\frac{m_t}{\upsilon}H{\overline{t}}(a_t+ib_t\gamma_5)t \ ,\]
where $m_t$ is the top-quark mass, $\upsilon$(= 174 GeV) the vacuum expectation value, and the coefficient a (b) denotes the CP-even (CP-odd) coupling, respectively.   

	Examples of Feynman diagrams for considered tt and ttH processes are presented in Fig. 1.
\bec 
{\includegraphics[width=0.25\textwidth]{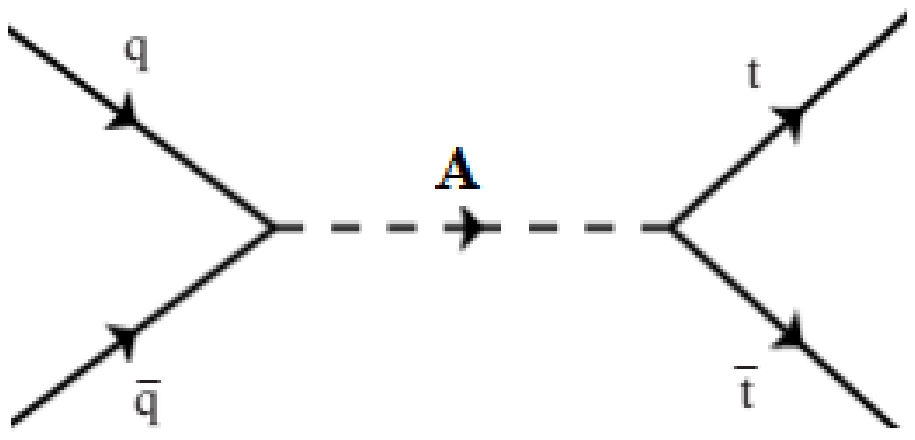}}\\
\bec
a)
\ec
{\includegraphics[width=0.59\textwidth]{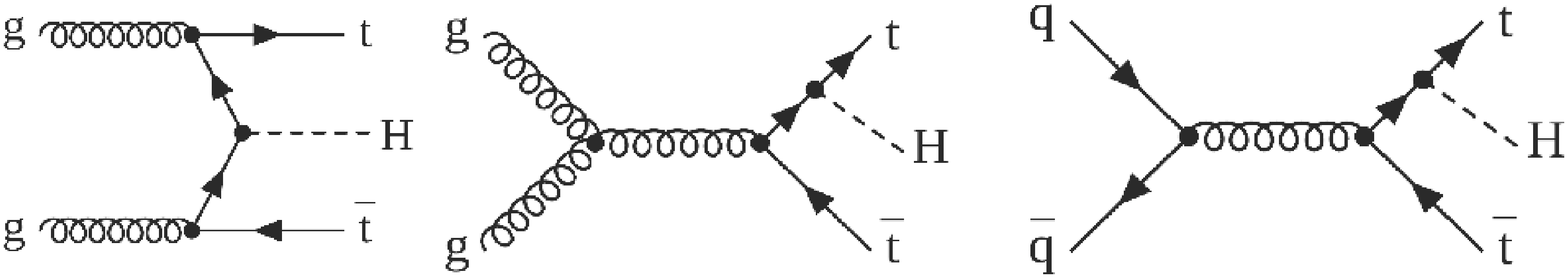}}\\
\bec
b)
\ec
\emph{{Fig.1.}} {\emph{Examples of Feynman diagrams for pp$\rightarrow$ A (a) and pp$\rightarrow$ ttH (b) production process, from \cite{4.}.}}
\ec    

It is necessary to reconstruct as many final particles as possible for disentanglement of the decay products of the exotic particles from the SM background. 
B-tagging identification, connected with b-quark signatures,
has following features and benefits for experimental determination of primary particles:
 
    $\bullet$ hadrons containing b quarks have sufficient lifetime;
    
    $\bullet$ presence of secondary vertex (SV);
    
    $\bullet$ tracks with large impact parameter (IP);
    
    $\bullet$ the bottom quark is much more massive, with mass about 5 GeV, and thus its decay products have higher transverse momentum; 
    
    $\bullet$ b-jets have higher multiplicities and invariant masses;
    
    $\bullet$ The B-decay produces often leptons. 

	We carried out a comprehensive computer modeling of MSSM Higgs boson mass using MadGraph, Pythia and Delphes programs. With the help of the program MadGraph, we carried out a calculation of the production cross sections of the processes under consideration. Simulation of further developments, i.e. all information on decomposition products and their kinematic data was produced using the Pythia program. For our calculations with Pythia we used the latest experimental constraints on the low tan$\beta$ region covered by ttH, H $\rightarrow$ tt processes, \cite{5.}. The response of the detector to the resulting array of events was carried out using Delphes program. We made a selection of events on the basis of additional kinematic restrictions associated with the peculiarities of the reactions under consideration and the b-tagging method.
	
	Let’s consider these processes separately and in more detail.
	\subsection{pp$\rightarrow$A$\rightarrow$tt process}
	The importance of the formation of a top quark pair is associated both with the possibility of good identification of top quarks using the b-tagging algorithm, and with the search for new physics due to the Yukawa constants of the top-quark and Higgs boson interaction, \cite{6.}. The SM, makes predictions for the coupling of the Higgs boson to top quark. Therefore, measurement of the decay rates of the observed state yields information which can be used to probe whether data are compatible with the SM predictions for the Higgs boson. Loop-induced vertices allow probing for BSM contributions of new particles in the loops. In addition, it must be said that measuring of the properties of top pair quarks also sheds light on the stability of the electroweak vacuum, \cite{7.}. The importance of this section is connected with improvement of the searches for the H $\rightarrow$ tt by studying the fully leptonic and the semi-leptonic final states, \cite{8.}. 
	According to our calculations presented in  \cite{9.}, presented in Fig.2. 
\bec 
{\includegraphics[width=0.29\textwidth]{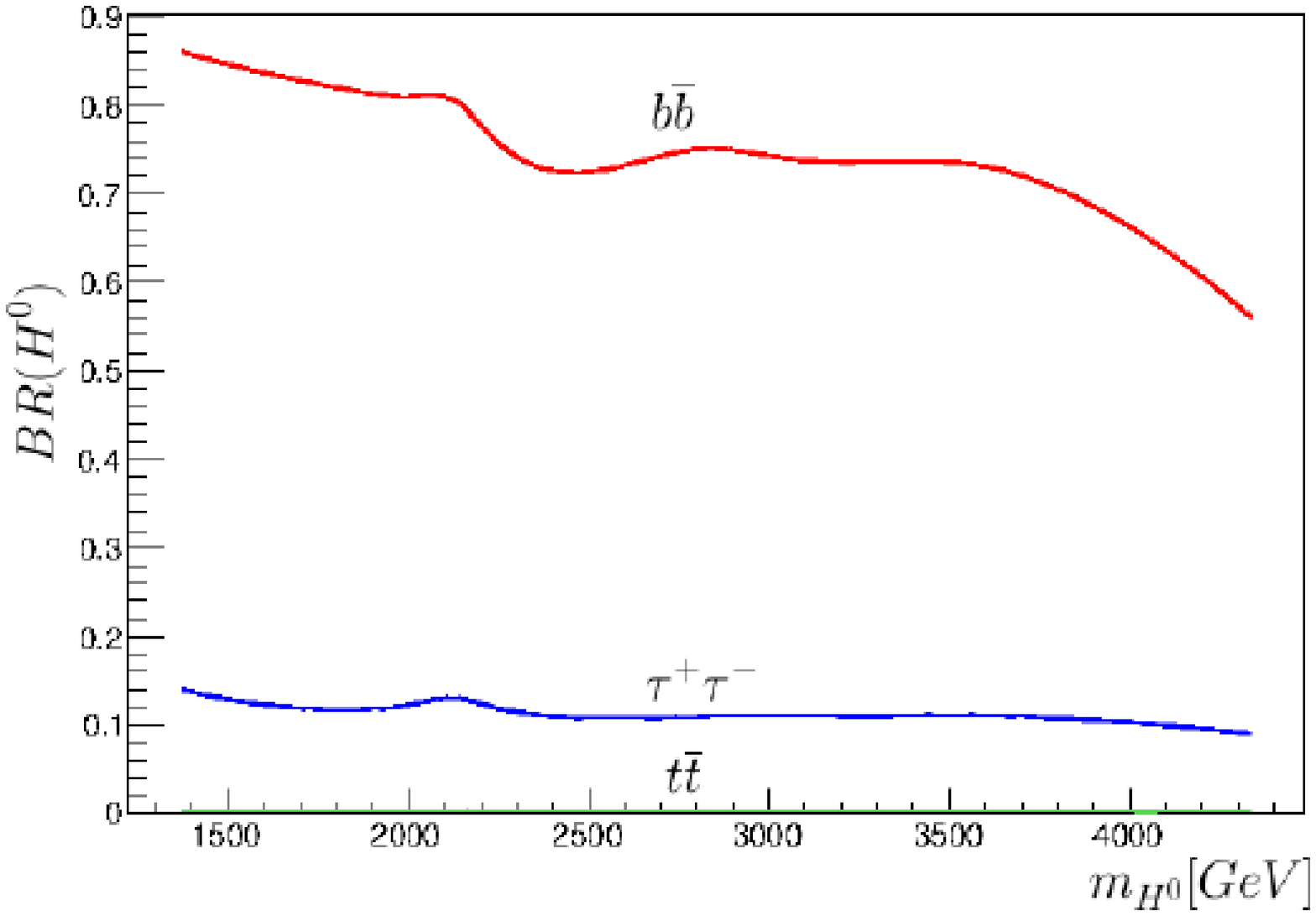}}\\
\emph{{Fig.2.}} {\emph{The branching fractions of H to bb (red), $\tau^{+}\tau^{-}$ (blue), tt (green).}}
\ec
the most probable decay channels for CP-even boson, H are the following: \\
    $\bullet$ bb;\\
     $\bullet$ $\tau^{+}\tau^{-}$;\\
      $\bullet$ $t\overline{t}$.
      
As we are dealing with massive MSSM particles, they prefer to decay into the most massive decay products, for example, into top-anti-top quark pair. So, we will consider the decay of CP-odd Higgs boson into top-anti-top quark pair, 
A$\rightarrow t \overline{t}$. With the help of the program MadGraph we calculated production cross section of the pp$\rightarrow$A$\rightarrow t\overline{t}$ process, presented in Fig. 3      
\bec 
{\includegraphics[width=0.32\textwidth]{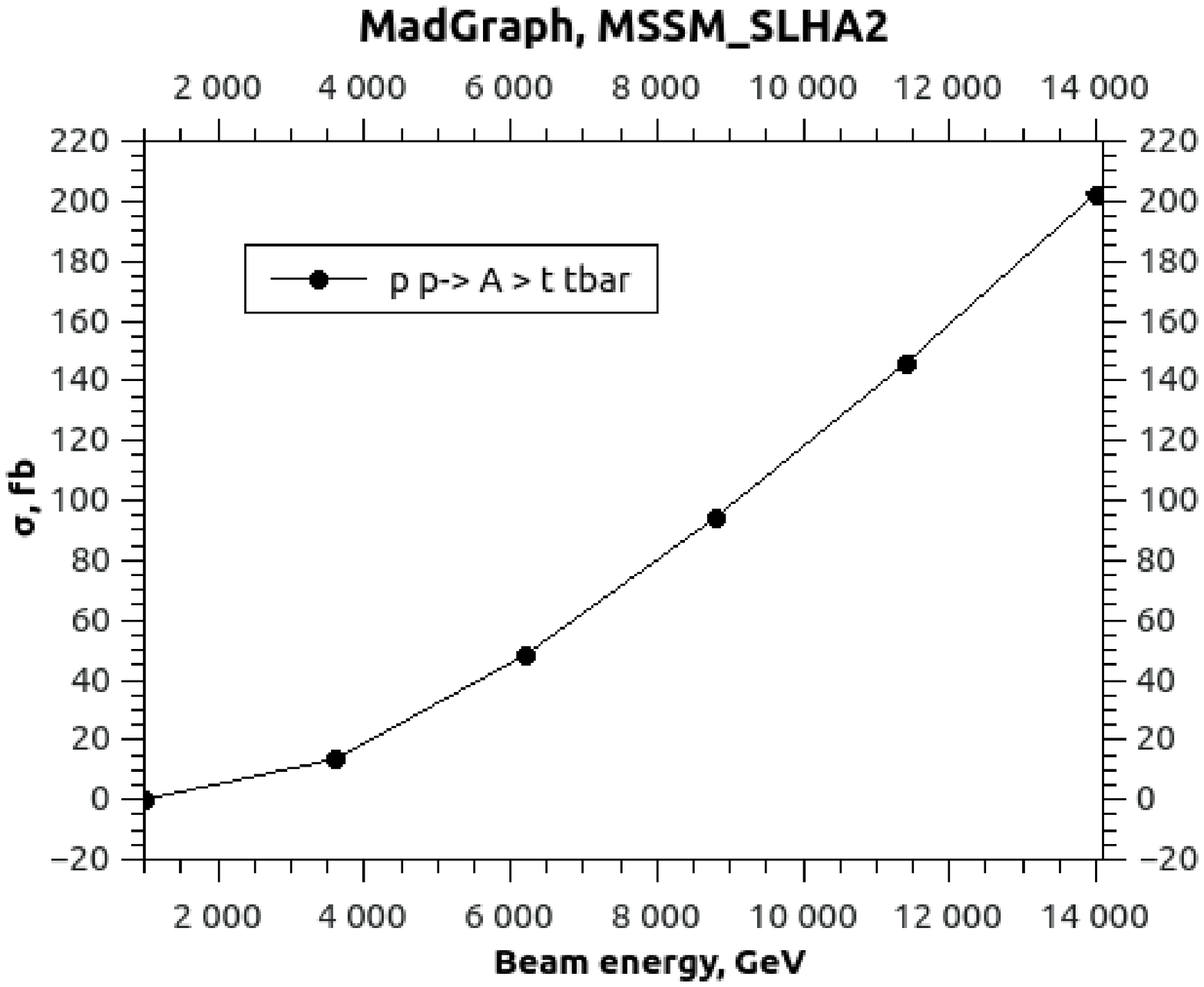}}\\
\emph{{Fig.3.}} {\emph{Production cross section of the pp$\rightarrow$A$\rightarrow t\overline{t}$ process.}}
\ec      
The increase of the production cross section with the energy at the LHC and its large value for the formation of A boson, about 0.2 pb at the energy of 14 TeV, leads to the conclusion about the importance of the consideration of this channel of formation and decay of the MSSM Higgs boson. Kinematic properties of decay products of A boson at the energy 14  TeV were modelled and presented in Fig. 4
\bec 
{\includegraphics[width=0.59\textwidth]{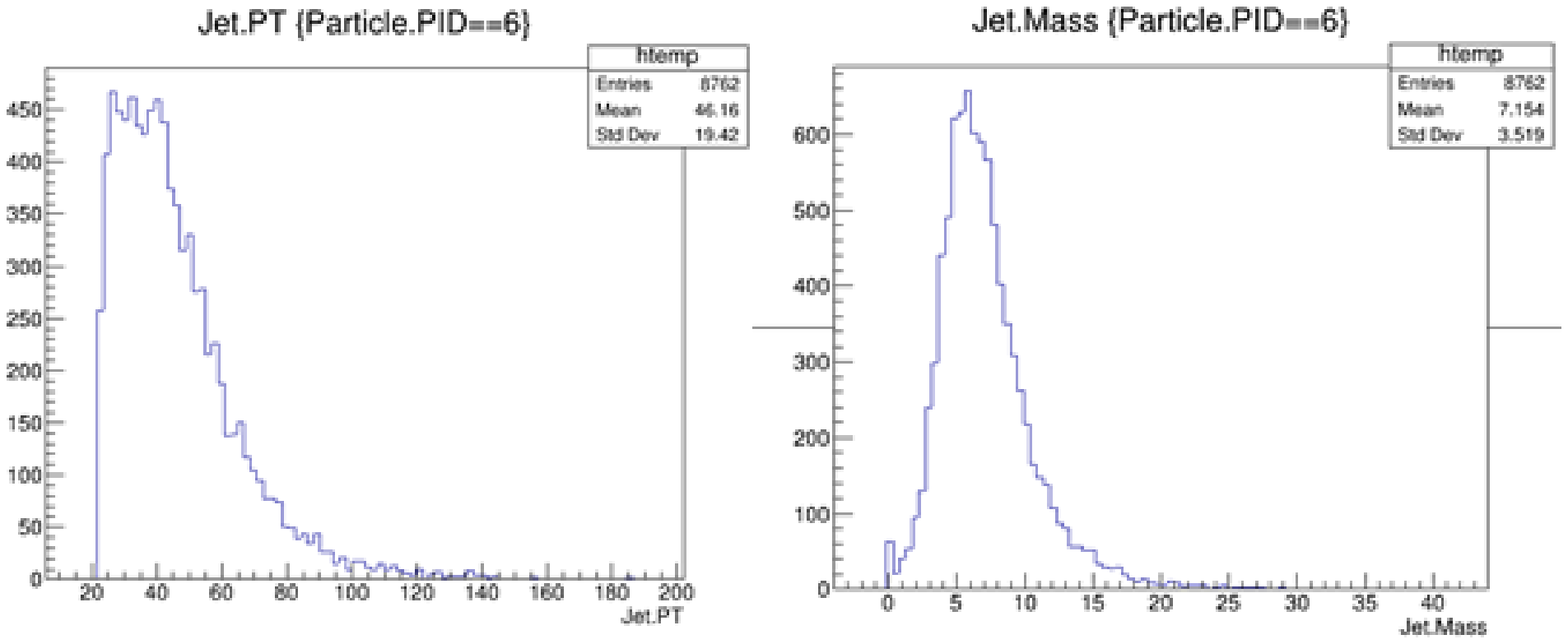}}\\
\bec
a)
\ec
{\includegraphics[width=0.29\textwidth]{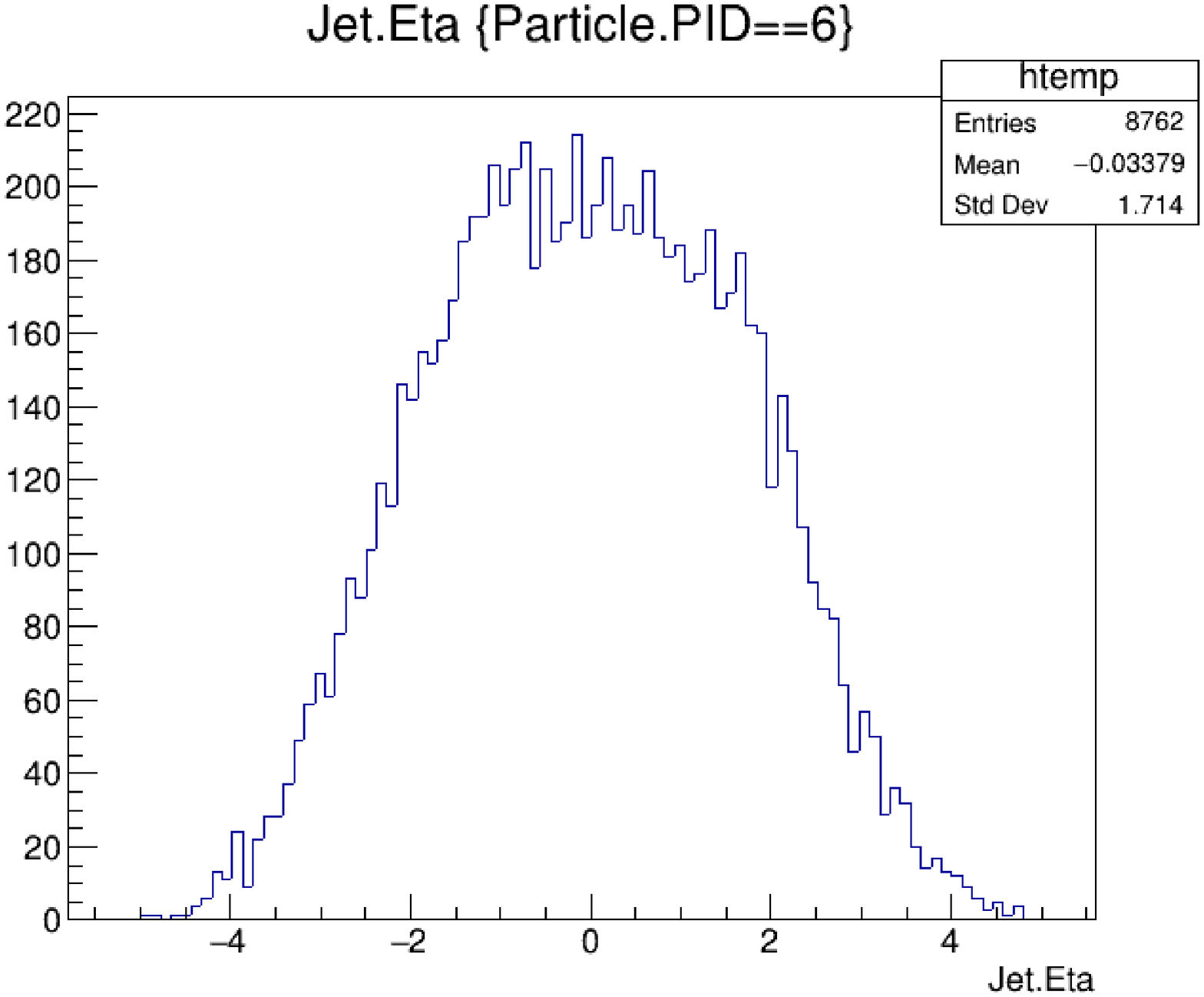}}\\
\bec
b)
\ec
\emph{{Fig.4.}} {\emph{Modeling of kinematic properties of jets from the reaction pp$\rightarrow$A$\rightarrow$tt : a) jet p$_T$ distribution (left) and jet mass (right); b) jet eta distribution. }}
\ec  
From Fig.4 we see that jet p$_T$  is maximal in the region of 30-50 GeV/c and then sharply decreases in the region of 120-140 GeV/c. Average jet mass is about 5-7 GeV/c, which is in accordance with the mass of the b-quark, into which the top quark decays with a probability of 99.8\%. The angular distribution of the decay products shown in Fig. 4 b) indicates the predominant direction of the decay products in the direction of angles from 35$^0$ to 90$^0$ to the axis of proton-proton collision. In Fig. 5 is presented distribution for jets in the momenta and angles.
\bec 
{\includegraphics[width=0.35\textwidth]{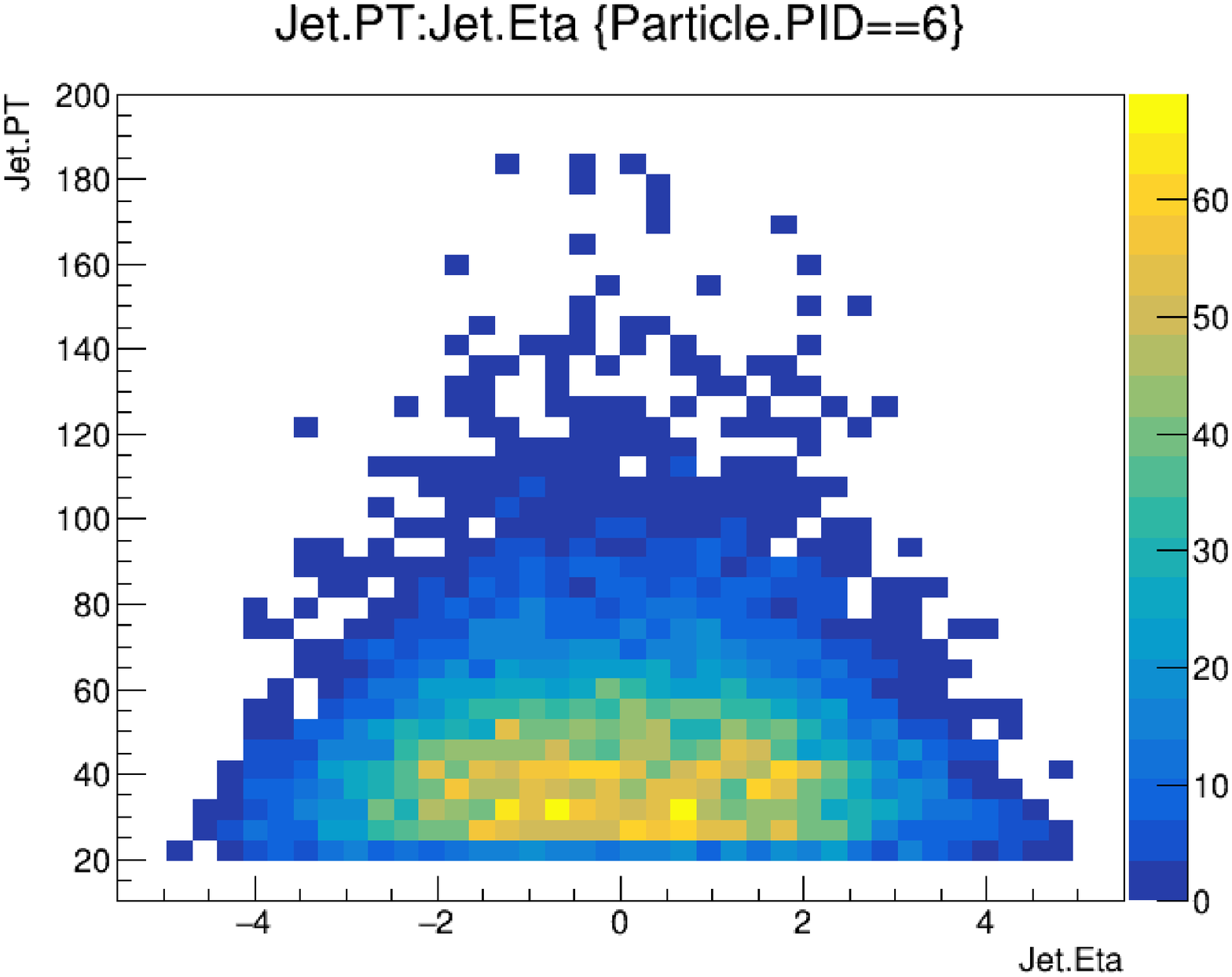}}\\
\emph{{Fig.5.}} {\emph{Distribution for jets in the momenta and angles for the reaction pp$\rightarrow$A$\rightarrow$tt .}}
\ec      
Using the distribution of Fig. 5 we can pick out the most high- energetic jets and present their separation in Fig. 6.
\bec 
{\includegraphics[width=0.35\textwidth]{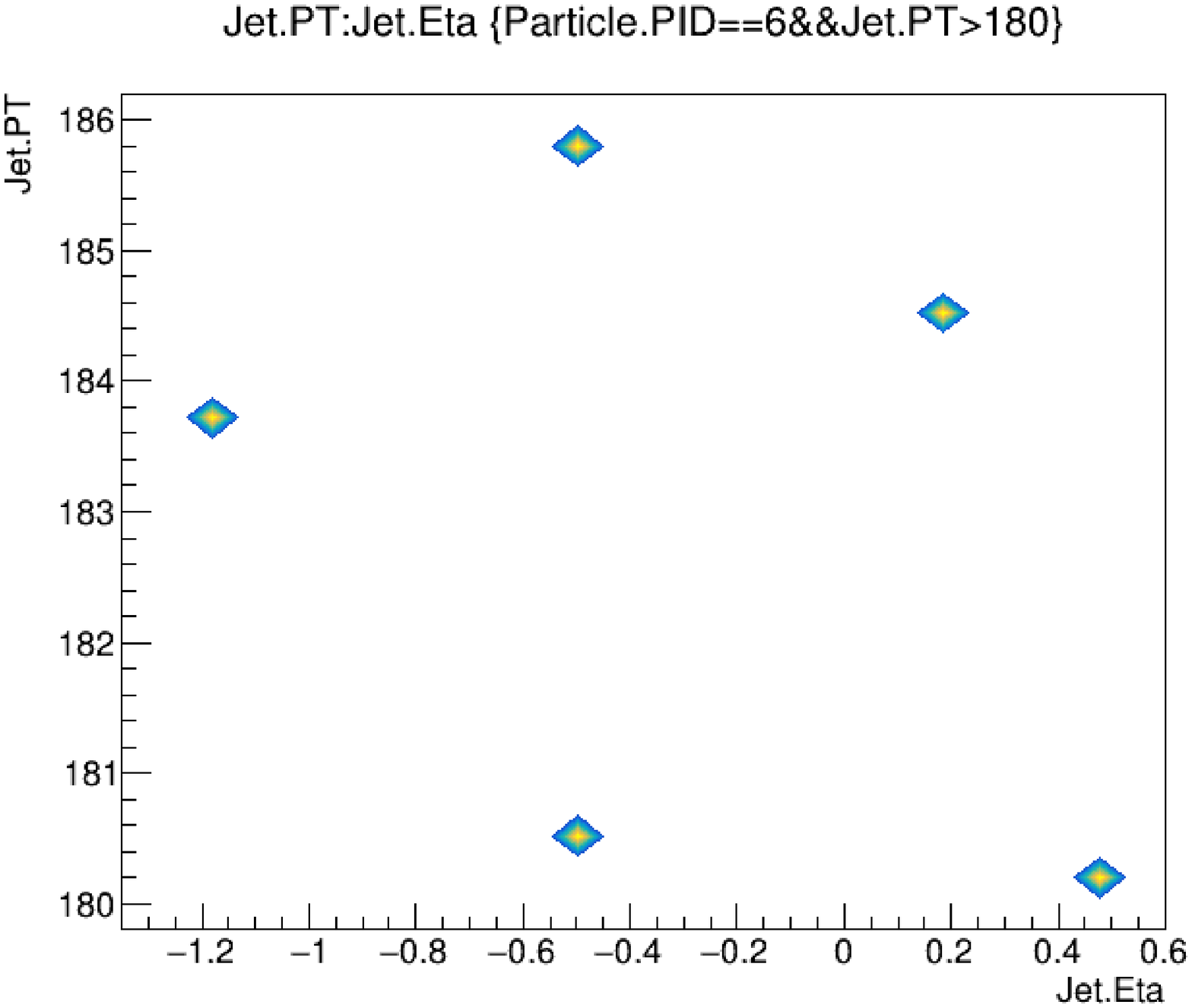}}\\
\emph{{Fig.6.}} {\emph{The most energetic jets in the p$_T$ range 180-186 GeV/c and jet $\eta > |1.2|$ for the reaction pp$\rightarrow$A$\rightarrow$tt .}}
\ec 
Using the data of Fig. 6 we can predict the mass of the A boson, which is about 360 GeV/c, since at high energies the momentum is equal to mass.
\subsection{ttH production process}
We considered combined analysis of proton-proton collision data at center-of-mass energies of $\sqrt{s}$ = 7, 8, and 13 TeV, corresponding to integrated luminosities of up to 5.1, 19.7, and 35.9 fb$^{-1}$, respectively.  In this experiment the observation of ttH production with a significance of 5.2 standard deviations above the background-only hypothesis, at a Higgs boson mass of 125.09 GeV was reported in \cite{4.}. Then we considered the decay process of the Higgs boson, H$\rightarrow$bb as the most probable decay process, \cite{9.}.

	Using the program MadGraph we calculated production cross sections pp$\rightarrow$Htt of Higgs boson via proton - proton interaction. Our calculations in the range of 2-14 TeV at the LHC are presented in Fig.7
\bec 
{\includegraphics[width=0.35\textwidth]{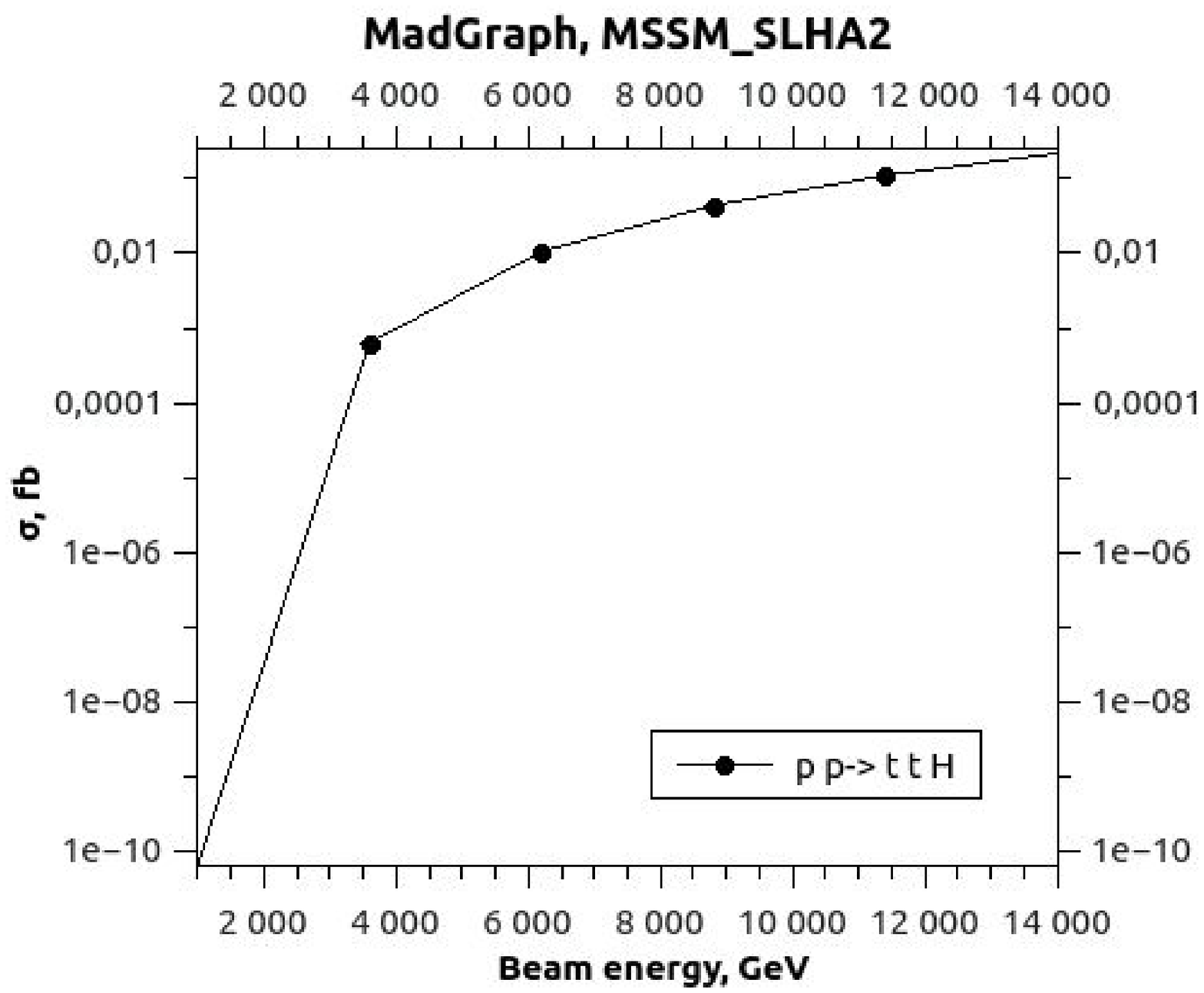}}\\
\emph{{Fig.7.}} {\emph{Production cross section of  pp$\rightarrow$Htt process\\ as the function of energy range at the LHC.}}
\ec	
With the program Pythia we simulated further development of events. The detector response to the received array of events was modeled by the program Delphes. Thus, our simulation was maximally close to the experimental conditions. 

	The results of calculations of jet mass range and eta distribution of jet are presented in Fig. 8. The events were selected with the following cuts: number of jets, N$_{charged} > $ or $\sim$ 4, transverse momentum, p$_T >$ 80 GeV, B$_{tag}$=1, mass of one b-jet, $M>4$ GeV. From jet distribution of Fig.4 we conclude that the mass of jets of about 16-20 GeV for minimal number of 4 jets corresponds to b-jet distribution. The corresponding angular distribution of jet flux signals about the selected distribution of the jet flow in the direction perpendicular to the proton collision axis with $\theta\sim$40$^0$-90$^0$. 
\bec 
{\includegraphics[width=0.39\textwidth]{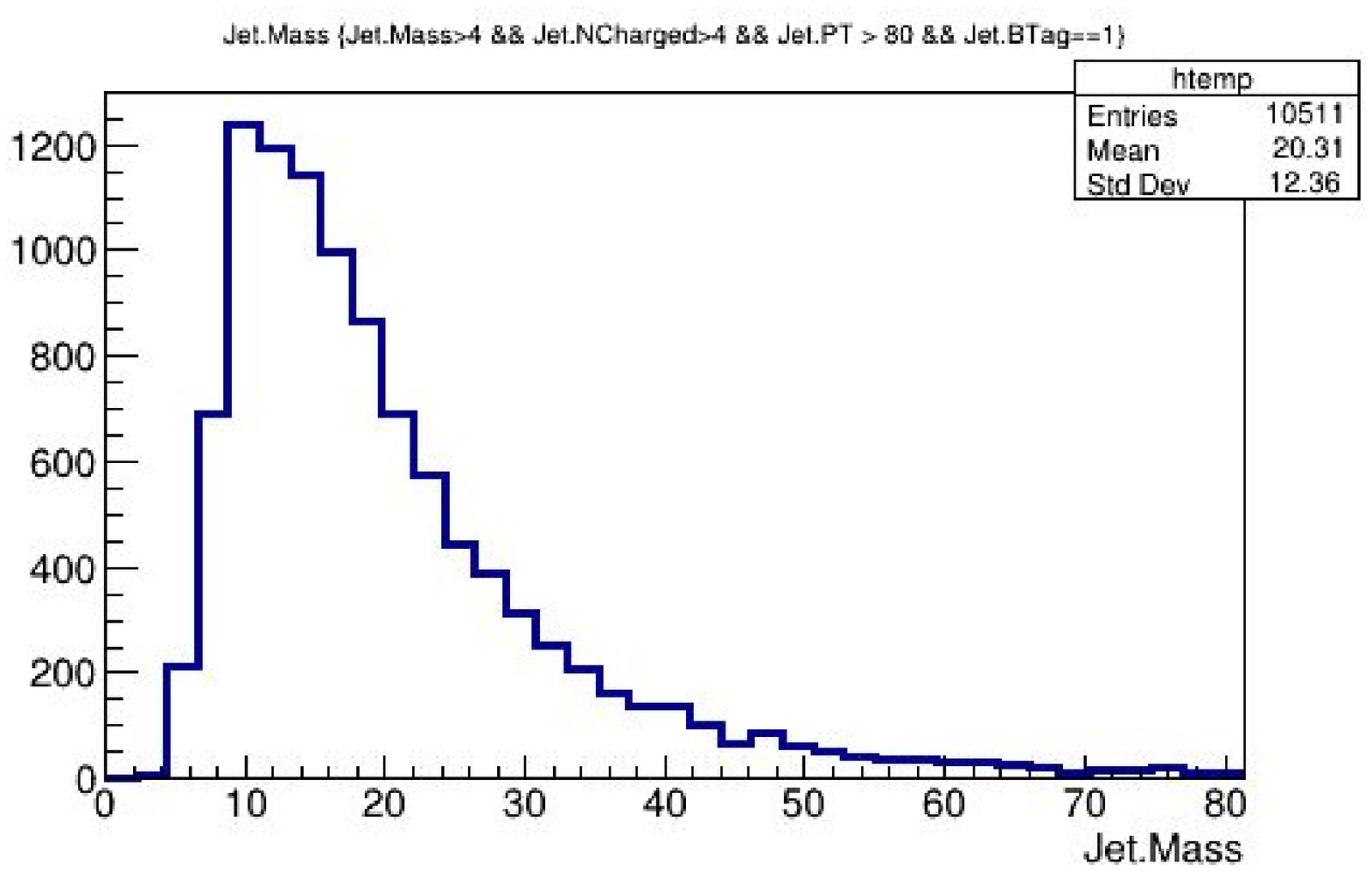}}\\
\bec
a)
\ec
{\includegraphics[width=0.39\textwidth]{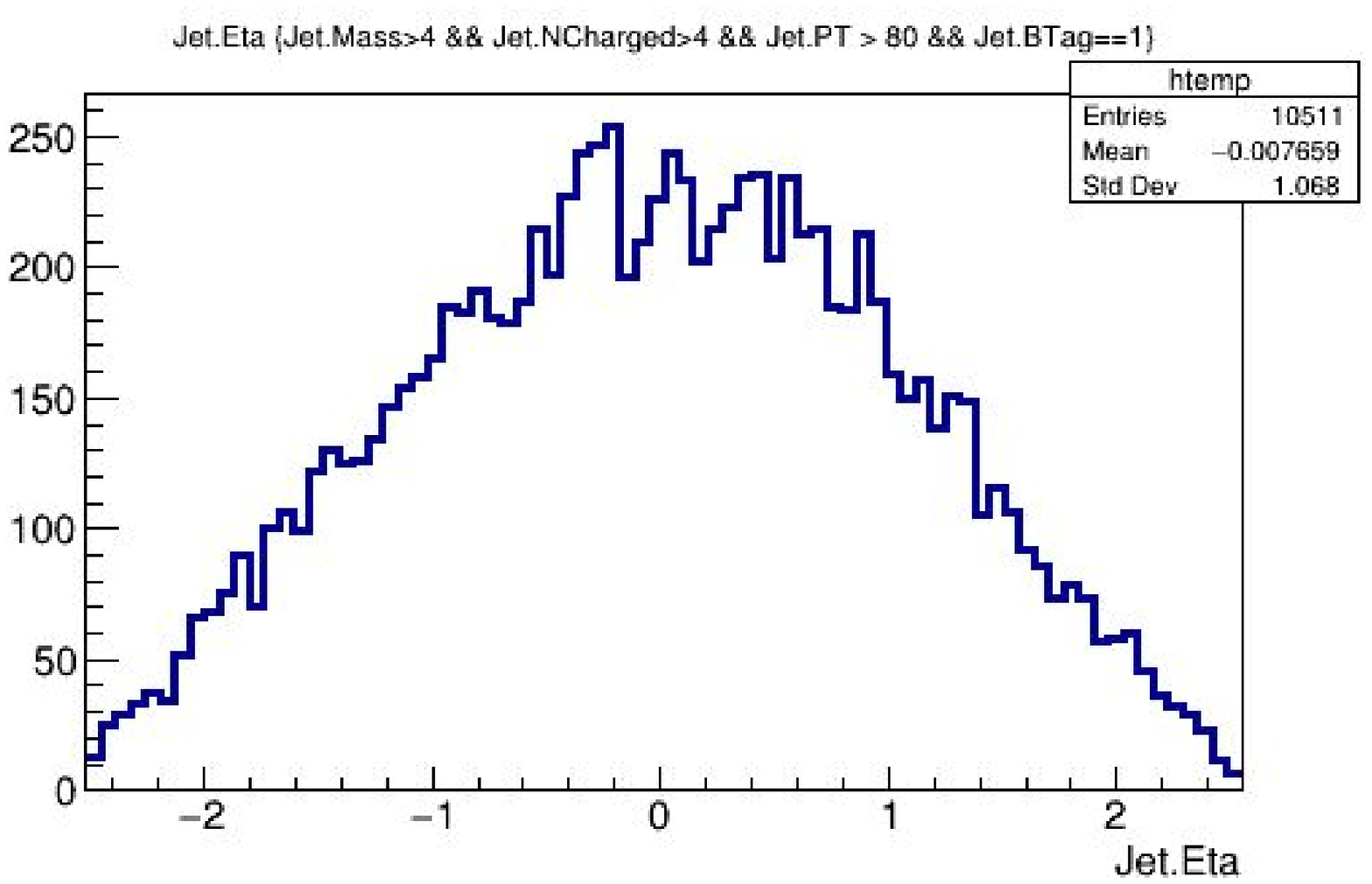}}\\
\bec
b)
\ec
\emph{{Fig.8.}} {\emph{Modelling of jet mass range (a) and angular distribution of jet (b). }}
\ec  	
As the result of detector response calculations for process pp$\rightarrow$Htt$\rightarrow$Hbbbb  with N=5000 initial events and corresponding cross section of about 0.517\ fb at 14 ТeV at the LHC we received the angular and p$_T$ jet distribution presented in Fig. 9. 	
\bec 
{\includegraphics[width=0.35\textwidth]{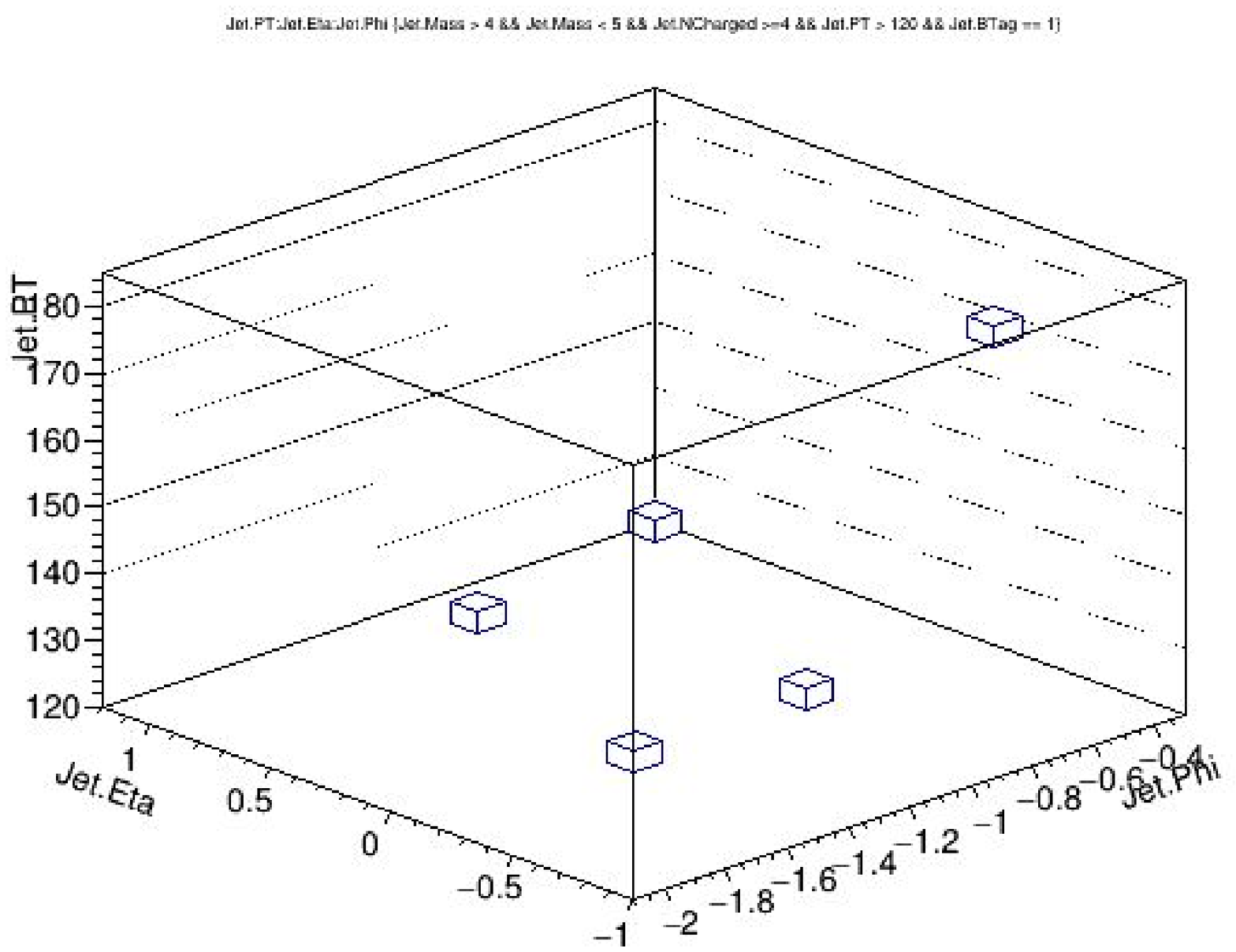}}\\
\emph{{Fig.9.}} {\emph{Modelled angular and p$_T$ jet distribution.}}
\ec
We have used tougher kinematic constraints: rapidity   $-0.5<y<0.5$, mass of  jets of about $4 <M<5$ GeV, number of charge jets, N$_{charged}>$4, transverse momentum, p$_T >$120 GeV, parameter of MSSM model, M$_H \sim$ 500 GeV. Thus, we selected the toughest and most massive jets that can be formed from the decay process of the CP-even Higgs boson of the MSSM model. As we can see from Fig. 9 the approximate mass of one jet is about 150-170 GeV/c . We used the fact that the protons each have an energy of 7 TeV, giving a total collision energy of 14 TeV. At this energy the protons move at about 0.999999990 c of speed of light, c. Knowing the most likely Higgs boson decay channel, H$\rightarrow$bb, we conclude that mass of CP-even Higgs boson is about  300-340 GeV/c. 

\section{Conclusion}
We have considered the most important channels of the MSSM Higgs boson production and decay. Since these channels are associated with the formation and decay of top quarks, whose properties shed light on the instability of an electroweak vacuum, the study of such reactions seems to us the most relevant. In addition, the MSSM Higgs bosons are the lightest supersymmetric particles predicted by supersymmetry. Therefore, finding their masses at the LHC collider is possible in the near future, which would remove a lot of theoretical questions related to symmetries and unification of interactions. Using programs MadGraph, Pythia and Delphes to simulate the processes and to model the response of the detector, as well as strict kinematic cuts on the angles and momenta of particles taken from the experimental data, we calculated the masses of A boson equal to 360 GeV/c and of H boson equal approximately to 320 GeV/c.


\label{page-last} 
\label{last-page}
\end{document}